\documentclass{PoS}

\title{Investigation of the overlap of excited bottomonium states with hybrid operators}

\ShortTitle{Investigation of the overlap of excited bottomonium states with hybrid operators}

\author{\speaker{C.~Ehmann}, T.~Burch and A.~Sch\"afer\\

        Institut f\"ur Theoretische Physik\\
        Universit\"at Regensburg\\
	D-93040 Regensburg, Germany.\\

        E-mail: \email{christian.ehmann@physik.uni-regensburg.de}\\
        }

\abstract{We analyze the overlap of color-octet meson operators with the $\Upsilon$ and the $\eta_b$ and their excited states,
especially the first radial excitations. Our analysis is based on NRQCD and 
includes all terms up to order $v^4$. We use a variety of source and sink operators as a basis for the variational method, 
which enables us to clearly separate the mass eigenstates and hence to extract the desired amplitudes. 
The results show the usefulness of the variational method for determining couplings to excited hadronic states.}

\FullConference{XXIVth International Symposium on Lattice Field Theory\\
                July 23-28, 2006\\
                Tucson, Arizona, USA}

\begin{document}
\section{Introduction}

We present a study of couplings to the excited states of bottomonium through 
the use of a variational method. 
There are two reasons for this endeavor:
We first wish to correct the interpretation given in Ref.\ \cite{luo} of 
there being no sign of low-lying excitations (e.g., radial) in heavy hybrid 
correlators by showing that they are indeed there after all, but that the coupling is simply too small to be accessible by the methods used in Ref.\ \cite{luo}. 
The point is that interpolators of any given quantum numbers will project out 
{\it all} states with those quantum numbers, regardless of whether the 
interpolators are {\it hybrid} in nature (i.e., containing gluonic fields; 
see, e.g., Ref.\ \cite{Burch}). 
Our second reason is to demonstrate the efficacy of the variational method 
for finding couplings of excited hadronic states. 
This has been recently studied for excited pions \cite{McNeile:2006qy}. 
Here, we show that ratios of couplings to a given state may be determined 
with good precision, even for resonance states and highly non-trivial currents.

\section{Method}
In the heavy quark limit we can readily define what a hybrid is: A hybrid excitation is a state which has large contributions of configurations where the quark-antiquark pair form a   color octet. Such a $q\bar{q}$ pair is accompanied by an arbitrary number of valence gluons to give a singlet. Since these states, where the $q\bar{q}$ pair is in a defined color representation, are not mass eigenstates of QCD, mixing between such configurations will occur. So we expect hybrid operators to have finite overlap with all excitations of a given state. 
We show this for the bottomonium system, where the mixing is actually 
suppressed compared to mesons containing lighter valence quarks.

To extract masses and couplings of excited states we rely on the variational method developed by 
Michael \cite{Mi85} and later refined by L\"uscher and Wolff \cite{LuWo90}.

In order to obtain the correlators, we first have to calculate the fermion propagators. Since bottom quarks are considered, the framework of NRQCD is applicable. We include all terms up to $O(v^4)$, where $v$ is the velocity of a quark, according to the power counting in \cite{Lepage:1992tx}.

The propagation of the fermions is given by:
\begin{equation}
\label{evolve}
\phi(\mathbf{y},t+a)=\left(1-\frac{aH_0(t+a)}{2n}\right)^nU_4^\dagger(t)\left(1-\frac{aH_0(t)}{2n}\right)^n\left(1-a\delta H(t)\right)\phi(\mathbf{x},t),
\end{equation}
where $H_0$ is
\begin{eqnarray}
H_0 & = & -\frac{\tilde{\Delta}}{2M}-\frac{a}{4n}\frac{\tilde{\Delta}^2}{4M^2}
\end{eqnarray}
and $\delta H$ is
\begin{eqnarray}
\label{rel_corr}
\delta H & = & -\frac{1}{8M^3}\tilde{\Delta}^2 \\ \nonumber
& & +\frac{ig}{8M^2}(\nabla\cdot\mathbf{E}-\mathbf{E}\cdot\nabla)-\frac{g}{8M^2}\mathbf{\sigma}\cdot(\tilde{\nabla}\times\mathbf{E}-\mathbf{E}\times\tilde{\nabla}) \\ \nonumber
& & -\frac{g}{2M}\mathbf{\sigma}\cdot\mathbf{B}.
\end{eqnarray}
The tildes denote improved versions of the corresponding derivatives. 
We use $n=2$, which is more than sufficient in our case. 
$\mathbf{B}$ and $\mathbf{E}$ are the magnetic and electric fields 
created via the usual clover formulation.

The last two terms of (\ref{rel_corr}) are responsible for the configuration mixing mentioned above.

We determine the quark mass for our simulation from finite momentum correlators for the $\Upsilon$ by tuning the kinetic mass extracted from the non-relativistic energy-momentum dependence to the experimental mass of the $\Upsilon$. Since we want to investigate the $\eta_b$ and the $\Upsilon$, we use pseudoscalar and vector currents. For both we have a normal and a hybrid version. Table \ref{optab} gives an overview of the local operators we use. The P-wave states are only needed to set the scale. In order to assemble our basis with more linearly independent operators, we additionally smear the quark and the antiquark field independently with two different smearing levels. 
In total twelve different operators, each at the source and the sink, are available for constructing the cross correlator matrix.

\vspace{0cm}
\begin{table}[!h!]
\begin{center}
 \begin{tabular}{|c|c|c|c|}
 \hline
 state & $J^{PC}$ & normal operator & hybrid operator \\
 \hline
 $\eta_b$ & $0^{-+}$ & $\chi^{\dagger}\phi$ & $\chi^{\dagger}\sigma_i B_i \phi$  \\
 $\Upsilon$ & $1^{--}$ & $\chi^{\dagger}\sigma_i\phi$ & $\chi^{\dagger}B_i \phi$  \\
 $\chi_{b0}$ & $0^{++}$ & $\chi^{\dagger}\sigma_iD_i\phi$ & - \\
 $\chi_{b1}$ & $1^{++}$ & $\chi^{\dagger}\epsilon_{ijk}\sigma_jD_k\phi$ & - \\
 $\chi_{b2}$ & $2^{++}$ & $\chi^{\dagger}(\sigma_iD_j+\sigma_jD_i-\frac{2}{3}\delta_{ij}\sigma_kD_k)\phi$ & - \\
\hline 
\end{tabular}
\label{optab}
\caption{Overview of the used operators $\hat O_i$. The hybrid versions of the P-waves are not needed.} 
\end{center}
\end{table}

We note in passing that the hybrid operators we use are related to higher twist contributions to the second moment of parton distributions (see \cite{Gockeler:2005jz} for a related topic): 
\begin{equation}
\bar{\psi}\epsilon_{ijk}F_{jk}\psi \quad \sim \quad \bar{\psi}\epsilon_{ijk}[D_{j},D_{k}]\psi.
\end{equation}
We do not explicitly calculate such a contribution (for bottomonium such a 
quantity is probably of little interest), we just point out the 
usually inherent difficulty behind what it is that we are after, especially 
for the excited states.

We determine the correlator matrix
\begin{eqnarray}
C_{ij}(t) & =& \langle \hat O_i(t) \hat{\overline O}_j(0) \rangle
\end{eqnarray}
and solve the generalized eigenvalue problem: 
\begin{eqnarray}
C_{ij}(t)\, \psi^n_j  & = & \lambda_n(t,t_0)\, C_{ij}(t_0) \psi^n_j ,
\end{eqnarray}
where $\lambda_n$ is the eigenvalue corresponding to the eigenvector $\vec\psi^n$.

Since due to fluctuations $C_{ij}$ is not exactly symmetric (although it is within errors), we symmetrize it by hand in order to make the diagonalization procedure more stable.

The eigenvalues are given by \cite{Mi85,LuWo90}
\begin{eqnarray}
\lambda_n(t,t_0) & = & A_n \, e^{-M_n(t-t_0)}[1+O(e^{-\Delta M_n(t-t_0)})],
\end{eqnarray} 
where $M_n$ denotes the mass of the $n$th state and $\Delta M_n$ the mass difference to the next state. 
So for large enough values of $t$, we have a single mass state in each channel.

The eigenvectors shed light on the couplings of the states to the operators in use. 
Again, this should hold for large values of $t$. 
That is easy to see by regarding the following relation, where the eigenvalues are reconstructed by applying the matrix on the eigenvectors: 
\begin{equation}
\frac{C_{ij}(t)\psi^n_j}{C_{kl}(t)\psi^n_l} \quad = \quad 
\frac{\langle O_i | \psi_j^n\overline O_j \rangle}{\langle O_k | \psi_l^n\overline O_l \rangle} \quad = \quad 
\frac{\langle O_i|n\rangle}{\langle O_k|n\rangle},
\end{equation}
where $\langle O_i|n\rangle$ stands for the overlap of the $i$th operator with the $n$th eigenstate. Consequently, we are able to make statements about the ratios of couplings of different operators to the same state.

\section{Results}
We are working on configurations provided by the MILC-collaboration \cite{Aubin:2004wf}. They were generated using improved staggered fermions and the L\"uscher-Weisz gauge. Table {\ref{configs} shows the parameters of the lattices used. For the lattice spacing there are two values given. The first one comes from the analysis of the spin-averaged 1P-1S splitting, the second one is given by the MILC-collaboration.

\vspace{0.5cm}
\begin{table}[h!]
\begin{center}
 \begin{tabular}{|c|c|c|c|c|c|}
 \hline
$\beta$ & $a^{-1}[MeV]$ & volume & $N_f$ & $am_{sea}$ & $am_{b}$ \\
\hline
8.40 & 2378/2279 & $28^3\times96$ & 0 & $\infty$ & 1.7, 1.8 \\
7.09 & 2097/2252& $28^3\times96$ & 2+1 & 0.0062/0.031 & 1.7, 1.8 \\
6.76 & 1495/1587& $20^3\times64$ & 2+1 & 0.01/0.05 & 2.4, 2.5 \\
\hline
\end{tabular}
\label{configs}
\end{center}
\caption{The three different lattices we use in our simulations together with the corresponding parameters. The first value for the inverse lattice spacing comes from the 1P-1S splitting, the second one is given in \cite{Aubin:2004wf}.}
\end{table}

\vspace{0cm}
\begin{figure}[h!]
\begin{center}
\resizebox{300pt}{!}{\includegraphics[clip]{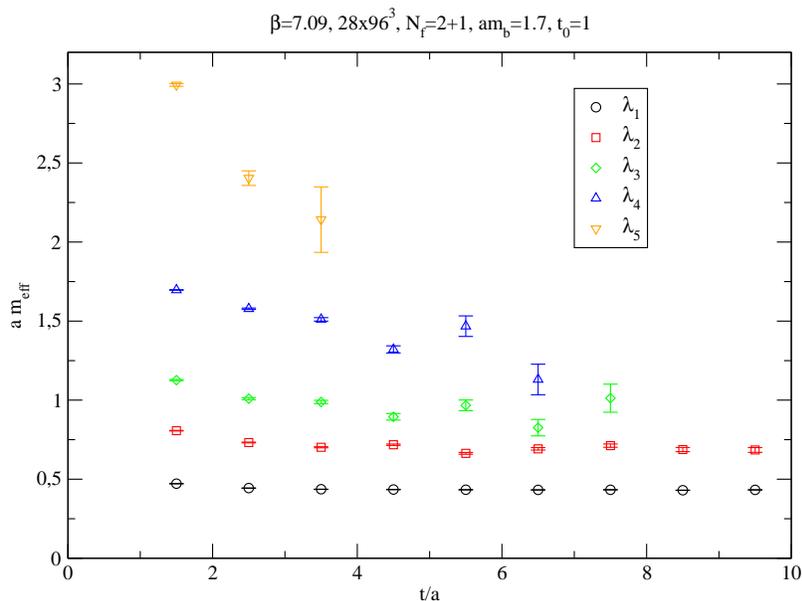}}
\end{center}
\caption{Effective masses of the five eigenvalues in the basis Nll(1), Nln(2), Nnn(3), Nww(4), Hll(5) for the dynamical lattice with $\beta=7.09$ and $am_b=1.7$. $t_0$ is the normalization timeslice.}
\label{ev1}
\end{figure}

Let us start with a five dimensional basis containing the operators Nll(1), Nln(2), Nnn(3), Nww(4), Hll(5). Here, the capital letter denotes the type of the operator (N=normal, H=hybrid), the lower case letters characterize the smearing of the quark and the antiquark (l=local, n=narrow, w=wide) and the number in brackets is the index of $\hat O_i$.

Figure \ref{ev1} shows the effective masses of the five eigenvalues of the $\Upsilon$ in this basis for the dynamical lattice with $\beta=7.09$ and $am_b=1.7$. The ratios of the couplings of the two local operators to the ground state and to the first radial excitation are plotted in Figure \ref{ratio1}. One can see that this ratio is about 1/90. This is somewhat suprising, since one would naively expect that radial excitations have a larger gluonic content than the ground state. About the same ratio is obtained also for all the other radial excitations. However, the scenario changes for the fifth state, which we identify with the ``hybrid excitation.'' In the left plot of Figure \ref{ratio2} one can see that the absolute value of the ratio of the couplings is about one.

An important check of our analysis is that the ratio of the amplitudes of the local operators should be independent of the number of extended operators contained in the basis. The right plot in Figure \ref{ratio2} shows that, in fact, this is the case.
The ratio is shown in the five dimensional basis given above and in the three dimensional basis Nll(1), Nnn(2), Hll(3). The same independence is also observed for all excited states. Therefore, we can conclude that the local operators are "approximately" orthogonal to the smeared ones. 
A similar sanity check also works for the large coupling ratio found for the 
hybrid excitation since it does not only appear as the highest state when we 
add more hybrid operators.

Since lattices with different spacings are accessible to us, we can search for possible scale dependencies. The ratios of the couplings for the ground and first excited state for two different lattice spacings are shown in Figure \ref{ratio3}. A clear dependence is visible. 
Given that the hybrid operator (in the numerator of the ratio) is not strictly a local current, but rather extended over a $2 \times 2$ clover, this seemingly strong scale dependence is not that surprising. 
Unfortunately, with only two different lattice spacings available (and the same extent of the hybrid operator in lattice units) and outstanding renormalization we cannot say anything more definitiv about this.

Similar results show up for the $\eta_b$, where the ratio for the non-hybrid states has changed to about 1/30, which can be traced back to the fact that in the $0^{-+}$ operator all three components of the $B$-field are included.

\begin{figure}[h!]
\begin{center}
\resizebox{300pt}{!}{\includegraphics[clip]{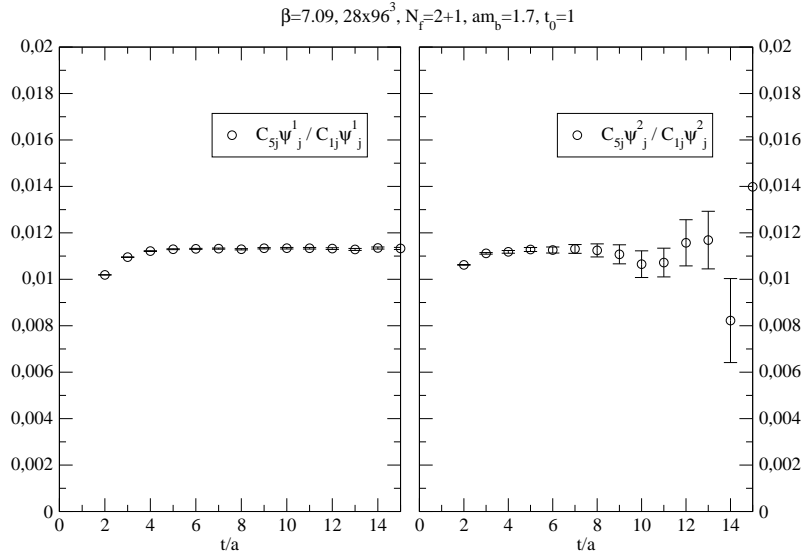}}
\end{center}
\caption{The ratio of the couplings of the two local operators to the ground state is shown in the left plot, the one to the first excitation in the right plot. In both cases we use the five dimensional basis of Fig.\ 1.}
\label{ratio1}
\end{figure}

 \begin{figure}[h!]
\begin{center}
\resizebox{320pt}{!}{\includegraphics[clip,angle=270]{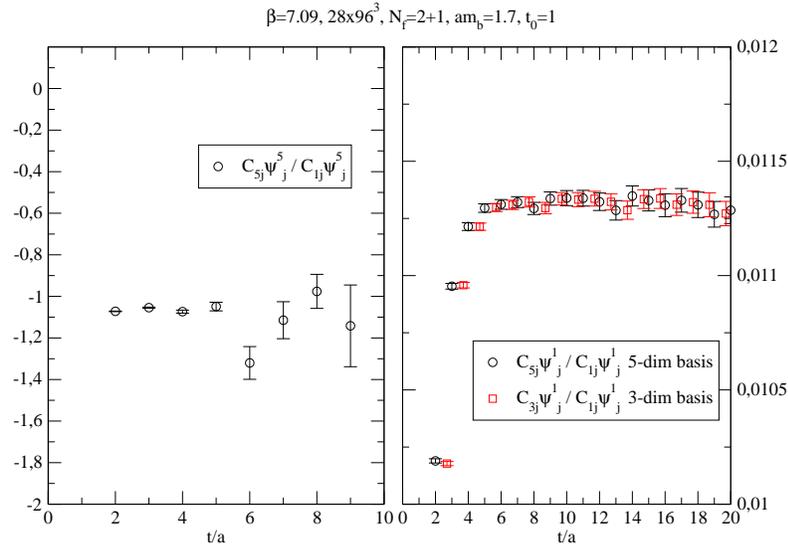}}
\end{center}
\caption{The ratio of the couplings of the two local operators to the hybrid excitation in the new basis is shown in the left plot. The right plot shows the ratio for the ground state in the three and five dimensional basis (the datapoints for the latter one are shifted by 0.3 in $t/a$ for the sake of clarity).}
\label{ratio2}
\end{figure}

\clearpage

\begin{figure}[h!]
\begin{center}
\resizebox{300pt}{!}{\includegraphics[clip]{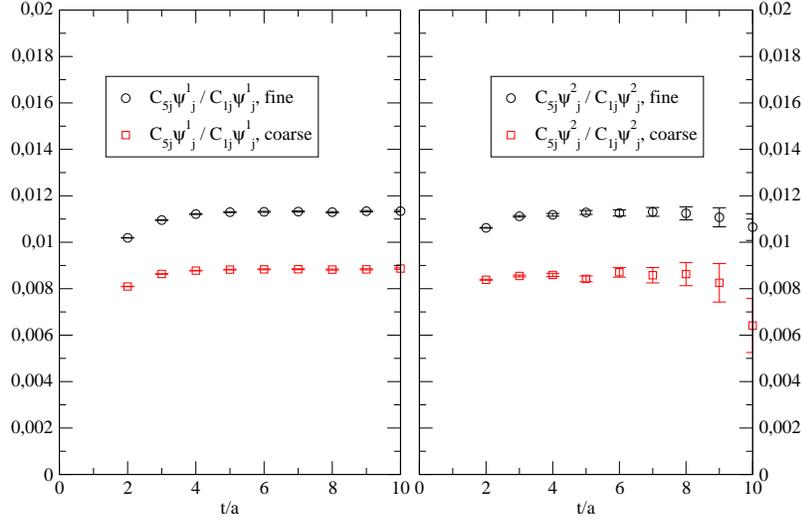}}
\end{center}
\caption{The ratio of the couplings of the two local operators to the ground state is shown in the left plot, the one for the first excitation on the right plot. Each plot is given for two different lattices with $a^{-1}=2101$ MeV (fine) and $a^{-1}=1497$ MeV (coarse).}
\label{ratio3}
\end{figure}

\acknowledgments
We would like to thank the MILC Collaboration for making their configurations publicly available.
This work is supported by GSI.

\end{document}